\documentclass[aps,prx,twocolumn,superscriptaddress,preprintnumbers]{revtex4-2}

\usepackage[T1]{fontenc}
\usepackage{amssymb}
\usepackage{graphicx}
\usepackage{color}
\usepackage{caption}
\usepackage{hyperref}
\usepackage{amsthm}
\usepackage{amsmath}
\usepackage{braket}
\usepackage{appendix}
\usepackage{subcaption}
\usepackage{comment}
\usepackage{bbold}
\usepackage{multirow}
\usepackage{booktabs}
\usepackage{ulem}
\usepackage{multirow}

\usepackage{tikz}
\usetikzlibrary{patterns}
\usetikzlibrary{arrows.meta,
                positioning}

\usepackage{physics}

\usepackage{float}

\usepackage{amsthm}
\usepackage{enumitem}
\usepackage{comment}

\begin{document}



\title{
Quantum Coherence of Topologically Frustrated Spin Chains
}

\author{S. B. Kožić}
\affiliation{Institut Ru\dj er Bo\v{s}kovi\'c, Bijeni\v{c}ka cesta 54, 10000 Zagreb, Croatia}

\author{G. Torre}
\affiliation{Institut Ru\dj er Bo\v{s}kovi\'c, Bijeni\v{c}ka cesta 54, 10000 Zagreb, Croatia}

\author{K. Delić}
\affiliation{Department of Mathematics, Faculty of Science, University of Zagreb, Bijeni\v{c}ka cesta 32, 10000 Zagreb, Croatia}

\author{F. Franchini}
\email{fabio@irb.hr}
\affiliation{Institut Ru\dj er Bo\v{s}kovi\'c, Bijeni\v{c}ka cesta 54, 10000 Zagreb, Croatia}

\author{S. M. Giampaolo}
\email{sgiampa@irb.hr}
\affiliation{Institut Ru\dj er Bo\v{s}kovi\'c, Bijeni\v{c}ka cesta 54, 10000 Zagreb, Croatia}

\preprint{RBI-ThPhys-2025-19}


\begin{abstract}

The study of entanglement and magic properties in topologically frustrated systems suggests that, in the thermodynamic limit, these quantities decompose into two distinct contributions.
One is determined by the specific nature of the model and its Hamiltonian, and another arises from topological frustration itself, resulting in being independent of the Hamiltonian's parameters.
In this work, we test the generality of this picture by investigating an additional quantum resource, namely quantum coherence, in two different models where topological frustration is induced through an appropriate choice of boundary conditions.
Our findings reveal a perfect analogy between the behavior of quantum coherence and that of other quantum resources, particularly magic, providing further evidence in support of the universality of this picture and the topological nature of this source of frustration.
\end{abstract}

\maketitle

\section{Introduction}

Although many-body quantum mechanics is known to exhibit a significantly richer phenomenology than its classical counterpart, the application of concepts borrowed from classical statistical mechanics has led to notable successes.
Thus, following Landau's approach, phases have been classified based on quantities such as correlation functions and local order parameters~\cite{1965193}.
However, the breadth of the phenomenology of many-body quantum systems is too broad to be fully described by such an approach.
A prominent example of a phenomenon that eludes such a classification is the existence of topological phases.
Such phases are characterized by discrete-valued nonlocal quantities that remain invariant under continuous deformations in the parameter space~\cite{PhysRevB.41.9377}. 
This phenomenon can be understood, for instance, in terms of the presence of one or more topological defects, which can be localized at the edges, such as the Majorana zero modes in the Kitaev wire~\cite{Kitaev_2001}, or can propagate freely in the bulk in the form of topological solitons, as in the fractional quantum Hall effect~\cite{wen1991, RevModPhys.71.S298}.

Since their discovery, several models have shown phenomenologies, often very different from each other, which can be encapsulated into the class of topological effects.
Among them, the topologically frustrated (TF) models have attracted considerable interest in recent years~\cite{Odavic2023,  catalano2025,PRXQuantum.5.030319,Rev1,Rev2, Rev3,Rev4,cinesi2,Dong2016}.
This phenomenon was first identified in one-dimensional antiferromagnetic spin chains (AFMs) with nearest-neighbor coupling and frustrated boundary conditions (FBCs)~\cite{Dong2016, Giampaolo_2019}. 
FBCs occur when periodic boundary conditions (PBCs) are imposed on a system with an odd number of spins.
This configuration breaks the periodicity of the AFM interaction, introducing a delocalized excitation into the system.
A similar phenomenology has also been identified in chains with even-number sites and tuning the frustration on and off by switching between periodic and anti-periodic boundary conditions~\cite{vicari}.
As a result of the presence of this excitation, the static~\cite{Marić_2020, fate2022} and dynamical~\cite{PhysRevB.105.184424} properties of TF systems acquire distinctive and universal behaviors that can be exploited for quantum technological applications~\cite{Rev2,PRXQuantum.5.030319,inprep}.
Moreover, this phenomenology extends far beyond the original class of models in which it was first discovered. 
A prominent example is the ANNNI model~\cite{SELKE1988213}, which features competing interactions between spins located at nearest and next-nearest neighbor sites along the lattice. 
In this system, topological frustration arises under periodic boundary conditions when the total number of spins is even but not divisible by four, leading to the appearance of two delocalized excitations~\cite{torre2024interplaylocalnonlocalfrustration}.

Among the various effects of imposing topological frustration on a one-dimensional system, one of the most surprising involves two fundamental quantum resources~\cite{RevModPhys.91.025001} such as the entanglement~\cite{entanglement_rev} and the non-stabilizerness (also known as magic)~\cite{PhysRevLett.128.050402}.
Specifically, in the frustrated phase~\cite{Giampaolo_2019, Torre2024, Odavic2023}, as the chain length diverges, the values of these resources can be decomposed into two distinct contributions.
The first is a somewhat {\it local} term, 
and stays unchanged even if the topological frustration is removed, i.e., when a different set of boundary conditions is applied to the system. 
In contrast, the second term has a topological nature and, therefore, remains independent of the specific Hamiltonian parameters. 
Thus, its value, which remains constant throughout the phase, can be evaluated near the classical point, namely the point at which the Hamiltonian reduces to a sum of mutually commuting terms, and the ground state can be calculated perturbatively.
This contribution to the quantum resources can be attributed to the presence of delocalized excitations induced by TF, which are known to be responsible for the topological nature of the TF phase, as proved through the Disconnected Entanglement entropy (DEE)~\cite{10.21468/SciPostPhysCore.7.3.050}.
A finite DEE is a unique signature of the long-range correlation typical of topological phases, although a characteristic topological invariant has not been identified yet for TF.

It is therefore natural to ask how general this picture is and whether it extends to other quantum resources.
Unfortunately, various quantum resources can have different structures with few points of contact, and thus it does not seem feasible to provide a general proof of this property or to perform a broad analysis. 
For this reason, in this paper we test the hypothesis of the generality of this picture by focusing on a further resource: Quantum Coherence (QC)~\cite{PhysRevLett.113.140401, RevModPhys.89.041003}.
The choice of the QC is not accidental but takes into account its relevance in several fields of research.
Indeed, QC plays a fundamental role across multiple areas of quantum physics. 
In quantum information, QC is a key resource for quantum computation and communication, enabling quantum speed-ups and secure protocols~\cite{PhysRevLett.113.140401, Streltsov_lett_2017}. 
In quantum thermodynamics, QC influences work extraction, quantum fluctuation theorems, and the fundamental limits of energy transfer in nanoscale systems~\cite{Engel2007,coherence_book, PRXQuantum.3.040323, lostaglio_description_2015}. 
In quantum many-body systems, its analysis can underlie the presence of quantum phase transitions, and the emergence of macroscopic quantum phenomena~\cite{Cirac2012, RevModPhys.89.041003, PhysRevB.108.125422, PhysRevA.110.022434,coherenceplenio}. 
Moreover, QC is essential for understanding the interplay between entanglement and non-classical correlations in open quantum systems~\cite{Levi_2014}. 

This paper presents an analysis based on a DMRG-driven numerical approach, which employs tensor networks to efficiently represent the ground state of the system. 
To construct a compact and accurate representation of quantum coherence (QC) as a univariate function, we also make use of the Tensor Cross Interpolation (TCI) algorithm~\cite{kožić2025computingquantumresourcesusing}. 
As is well established, QC is a basis-dependent quantum resource. 
Since it is unfeasible to analyze all possible bases in the Hilbert space of a large qubit system, our investigation is primarily conducted in the computational basis. 
This choice is motivated by two main reasons. 
First: near the classical point and in absence of topological frustration, the ground state approaches a GHZ-like form in this basis, where quantum coherence reaches a minimum. 
Second: prior studies on quantum magic in these models have also used the computational basis, allowing for a more direct comparison between the behavior of coherence and magic.
The results of our analysis are consistent with earlier findings on entanglement entropy and non-stabilizerness entropy, suggesting the possibility of a universal structure shared by entropy-based quantum resource measures.


The paper is structured as follows. In Sec.\ref{DMRG_TCI_Section}, we provide a brief overview of the numerical approach proposed in~\cite{kožić2025computingquantumresourcesusing} for computing the QC as measured by the Relative Entropy of Coherence (REC). 
In Sec.~\ref{Models}, we describe the models under consideration and summarize their known phenomenology. 
Section~\ref{Results} presents our results, demonstrating that, similar to entanglement and magic, for large systems, the REC in the computational basis decomposes into two distinct contributions: one is local, and the other is of topological origin. 
Finally, in Sec.~\ref{conclusion} we summarize the results and discuss future developments.
Additional important results are presented in the appendices. Appendix~\ref{ANNNI_coherence} contains the analytical computation of the REC in the computational basis for the ground state of the ANNNI chain near the classical point. 
In Appendix~\ref{rotation}, we present our numerical analysis considering other bases generated by rigid rotations and show that our findings, that is the decomposition of REC between a local and a topological contribution, are valid beyond the computational basis on which we concentrated in the bulk of the manuscript.
Lastly, Appendix~\ref{Bond} addresses the numerical challenges associated with these evaluations.

\section{Measuring Quantum Coherence}\label{DMRG_TCI_Section}

Over the years, several different measures of QC have been suggested~\cite{RevModPhys.89.041003}, each of them with its positive and negative sides.  
Since we are interested solely in pure states, in this work, we resort to the Relative Entropy of Coherence (REC)~\cite{PhysRevLett.113.140401}.
For a state with a density matrix $\rho$ the REC is defined as $ C(\rho) = S(\rho_{\textrm{diag}}) - S(\rho)$, where $S(\rho)$ stands for the von Neumann entropy and $\rho_{\textrm{diag}}$ is the matrix obtained by $\rho$ by deciding on a reference basis frame and setting all off-diagonal elements in this basis to zero. 
For a pure state, i.e., for $\rho=\ket{\psi}\!\bra{\psi}$, $S(\rho)$ vanishes and the REC reduces to
\begin{equation}\label{QC_pure}
    C(\rho)=-\sum_{i=1}^{2^L} c_i^2 \log_2 c_i^2.
\end{equation}
where $L$ is the number of qubits in the system, and $c_i$ are the coefficients of the state $\ket{\psi}$, in the base $\{\ket{i}\}_{i=1}^{2^L}$, i.e. $\ket{\psi}=\sum_{i=1}^{2^L} c_i \ket{i}$.
It is worth emphasizing that the expression for REC in \eqref{QC_pure} clearly shows that, like all other QC measures, this quantity depends on the chosen basis for calculation. 
Throughout this work, we will consistently evaluate REC on the computational basis.

Despite its apparent simplicity, the computation of Eq.~\eqref{QC_pure} is challenging, and, for large systems, it becomes unfeasible without an efficient representation of $\ket{g}$. 
But luckily this kind of representation exists and it is provided by the Matrix Product States (MPS)~\cite{orus_tensor_2019,biamonte2020lecturesquantumtensornetworks, PhysRevLett.69.2863,catarina_density_matrix_2023, ORUS2014117}.
Within such a representation, to evaluate the REC we observe that it is the sum of the components of the vector $f(\boldsymbol{c}^\intercal) = (f(c_1),\ldots, f(c_n))$, obtained by applying the function $f(x):= -x^2\log_2 x^2$ element-wise to the system's ground state. 
This sum can be evaluated by sampling the set $\{f(c_i)\}_{i=1}^{2^L}$ exploiting 
the family of tensor cross-interpolation (TCI) algorithms~\cite{fernández2024learningtensornetworkstensor, PhysRevX.12.041018} that allows us to escape the exponential scaling of the number of coefficients with the size of the system $L$ in~\eqref{QC_pure}~\cite{Savostyanov2011}.
It is worth noticing that, TCI algorithms are stable; within a given tolerance, the system consistently converges after a sufficient number of function evaluations for a specified bond dimension~\cite{fernández2024learningtensornetworkstensor}. 
To summarize, to evaluate the REC we follow a three-step algorithm~\cite{kožić2025computingquantumresourcesusing}. 
At first, we use a DMRG code to compute the ground state of the system. 
Hence, we exploit a TCI algorithm to sample the function $f$ of the ground state coefficients. 
Finally, to evaluate the REC, we contract the MPS resulting from the TCI decomposition.

\section{Models}\label{Models}

Using REC, we tested our hypothesis on quantum resources of two different one-dimensional models that can, under suitable conditions, show the presence of TF.
Both these models can be obtained for a particular choice of parameters, from the following Hamiltonian
\begin{equation}
    \label{eq:Hamiltonian}
    H=J_1\sum_{i=1}^L \sigma_{i}^z\sigma_{i+1}^z + J_2\sum_{i=1}^L \sigma_{i}^z\sigma_{i+2}^z +  h\sum_{i=1}^L\sigma_i^x.
\end{equation}
In Eq.~\eqref{eq:Hamiltonian} $\sigma_{i}^\alpha$ with $\alpha= x,\,y,\,z$ stand for the Pauli operators acting on the $i$-th spin of the system, $J_1$ and $J_2$ are, respectively, the next-neighbor and the next-to-nearest neighbor interactions, and $h$ is a local transverse field.
Unless stated otherwise, we assume periodic boundary conditions, i.e. $\sigma_{i}^\alpha\equiv\sigma_{i+L}^\alpha$.

Setting $J_2=0$ we recover the well-known Ising model with a transverse field~\cite{sachdev1999quantum, Franchini17} that shows a quantum phase transition at $|h|=h_c\equiv 1$. 
Choosing $J_1>0$ and imposing FBCs, for $0<|h|<h_c$, the system is in a topologically frustrated phase with a gapless energy spectrum in the thermodynamic limit. 
For $h=0$ (the {\it classical point} mentioned above), there are $2L$ exactly degenerate lowest energy states, namely the mutually orthogonal kink states $\ket{k^+}$ and $\ket{k^-}$, which are individually separable and incoherent in the computational basis.
Indeed, because of PBCs and $L$ being odd, it is impossible to have perfect Neel (staggered) order and therefore these states present a magnetic defect, i.e. two parallel aligned neighboring spins, which can be placed on any bond of the ring.
Hence $\ket{k^+}$ and $\ket{k^-}$ are the two states with the defect between the $k$-th and the $k+1$-th sites and differ from each other by the orientation of the defect.
Switching on a finite magnetic field (i.e., setting $h > 0$) leads to a hybridization of these states, thereby lifting their degeneracy and giving rise to the resulting (gapless) energy band.
In particular, in the limit $h\rightarrow0^+$ the GS is given by $\ket{g}=\frac{1}{\sqrt{2L}}\sum_{k=1}^L (\ket{k^+}+\ket{k^-})$ and the whole band can be characterized as a single delocalized kink excitation with some momentum.

On the contrary, considering $J_2>0$, we recover the so-called ANNNI model in which, independently of boundary conditions and the sign of $J_1$, there is always geometrical frustration induced by the competition between the two interactions.
The phase diagram of this model is rich, featuring four distinct phases that depend on the dominant interaction and the value of the magnetic field~\cite{SELKE1988213}.
Among these, one of the most peculiar is the so-called antiphase which can occur when $\kappa = \vert J_1 \vert / J_2 > 1/2$. 
In such a phase, TF can be induced considering a ring made of an even number of spins that is not an integer multiple of four ($L = 4n + 2, n \in \mathbb{N}$).
On the classical line $h=0$, the presence of TF increases the degeneracy of the ground state manifold, which goes from a dimension equal to 4 (in the absence of TF) to $L^2/2$ if $J_1\neq 0$ or to $L^2$ in the limiting case $J_1= 0$, where the chain decomposes into two independent subchains for the even and odd sites.
Turning on a small transverse field, the degeneracy is lifted and the system admits a unique ground state given in Eq.~\eqref{ANNNI_GS}, that can be described in terms of a distance-modulated superposition of kink states~\cite{torre2024interplaylocalnonlocalfrustration} (see Appendix~\ref{ANNNI_coherence} for details). 
Regardless of the presence or absence of the TF, by fixing the value of the ratio $\kappa$ and further increasing the value of $h$, the system undergoes two successive phase transitions. 
In the first, the system exits the antiphase and enters the so-called floating phase, whose nature has not yet been fully understood. 
It is characterized by a violation of the area law in terms of entanglement and is conjectured to be in a Luttinger liquid phase with algebraic correlations. 
Subsequently, the second transition marks the passage from the floating phase to a paramagnetic phase, where the dynamic is dominated by the transverse field.

\begin{figure}[t!]
    \centering
    \begin{subfigure}[t]{0.95\columnwidth}
        \centering        \includegraphics[width=0.95\columnwidth]{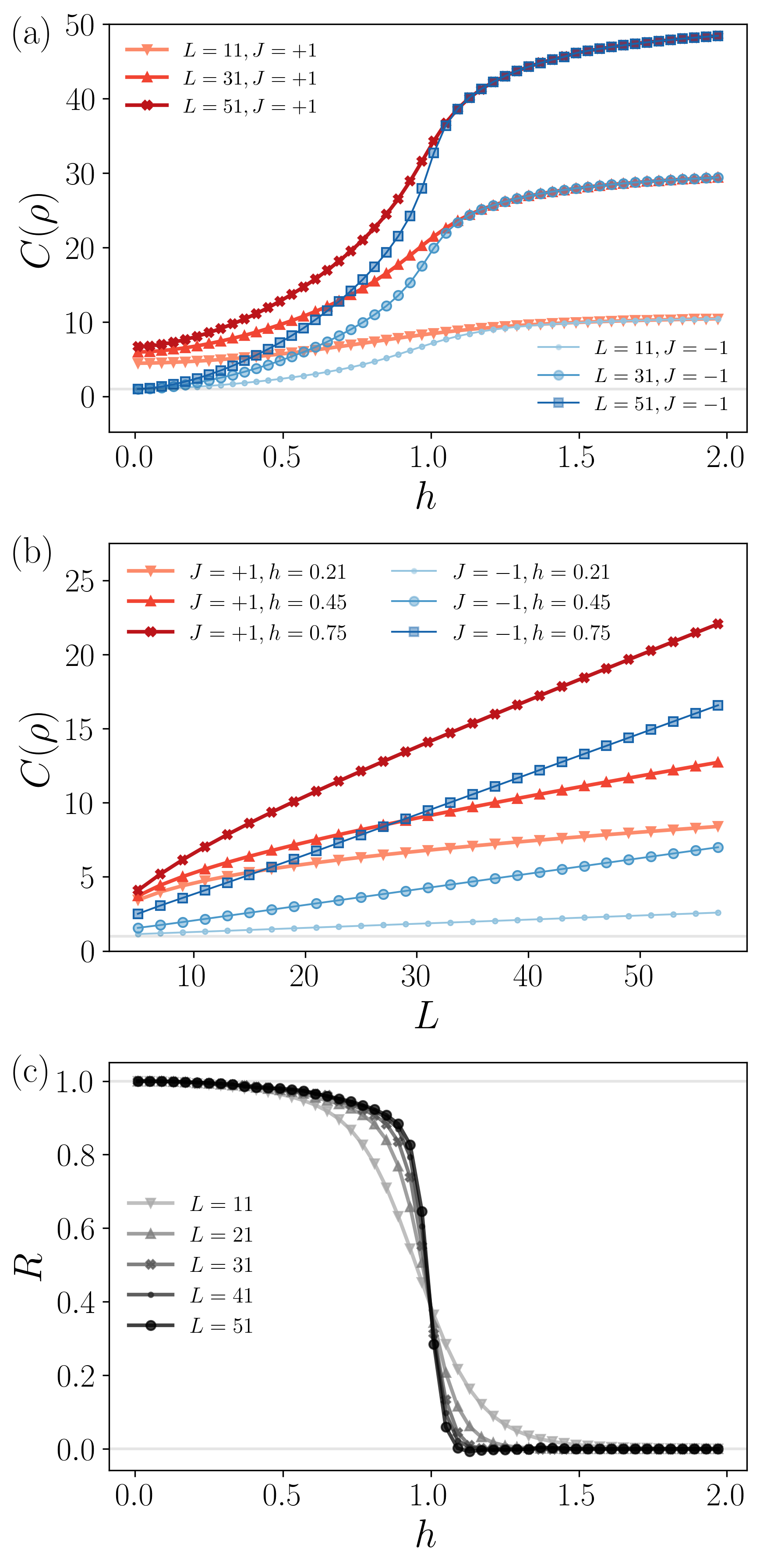}
    \end{subfigure}
    \caption{(a) Quantum Coherence of the ground state of the TF Ising chain (red full markers) and of the corresponding unfrustrated FM model (blue empty markers) as a function of the magnetic field and for different values of the system size. (b) The same quantities are plotted as a function of the system length for different values of the magnetic field. (c) The ratio $R$ in eq.~\eqref{eq:R_def} for the Ising model as a function of the magnetic field and for different system sizes.}
    \label{Ising_results}
\end{figure}

\section{Results}
\label{Results}

Having introduced the models under consideration and the TCI-based method for computing the REC, we now proceed to present and discuss our results. 
In this section, we focus specifically on evaluating the REC in the computational basis, while a generalization to a broader class of bases is provided in Appendix~\ref{rotation}.
The computational basis comprises states that are tensor products of the eigenstates of $\sigma^z_i$, where $i$ labels the sites of the system.
Accordingly, each basis state can be expressed as $\ket{\phi} = \bigotimes_{i=1}^N \ket{\phi_i}$, with each $\ket{\phi_i}$ equal to either $\ket{\uparrow_i}$ or $\ket{\downarrow_i}$.
Once the RECs have been evaluated, we test our hypothesis that, in the thermodynamic limit, any quantum resource computed for the ground state of a TF model decomposes into a sum of a local term and a topological term, the latter independent of the specific Hamiltonian parameters.
To this end, we employ the same approach previously used in~\cite{Torre2024} for Entanglement Entropy and in~\cite{Odavic2023,catalano2025} for the Stabilizer R\'{e}nyi Entropy.
This method involves analyzing the ratio
\begin{equation}
 \label{eq:R_def}
R=\frac{C(\rho_{\textrm{fr}})-C(\rho_{\textrm{unfr}})}{C(\rho_{\textrm{fr}}^{h\rightarrow 0^+})-C(\rho_{\textrm{unfr}}^{h\rightarrow 0^+})},
\end{equation}
where $\rho_{\textrm{fr}}$ and $\rho_{\textrm{unfr}}$ denote the ground state density matrices of the topologically frustrated and unfrustrated models, respectively, characterized by having the same correlation length.
According to our hypothesis, these two states share an identical local contribution to the QC in the thermodynamic limit, while differing in their topological terms.
Thus, subtracting one from the other isolates the difference between the topological contributions.
Similarly, $\rho_{\textrm{fr}}^{h\rightarrow 0^+}$ and $\rho_{\textrm{unfr}}^{h\rightarrow 0^+}$ correspond to the respective ground state density matrices near the classical limit, as $h \rightarrow 0^+$, where the Hamiltonian comprises mutually commuting terms.
If our hypothesis holds, the difference in coherence between these two states should likewise capture just the difference between the topological components.
As a result, the ratio defined in Eq.~\eqref{eq:R_def} compares two independent estimates of the same topological difference.
If this difference is indeed topological in nature, the ratio should remain invariant under variations of the system parameters, provided the system stays within the same macroscopic phase.
Therefore, if our hypothesis is valid—namely, that in the thermodynamic limit the QC of a topologically frustrated system splits into local and topological contributions—then the ratio in Eq.~\eqref{eq:R_def} should converge to 1 throughout the topologically ordered phase.

In Fig.\ref{Ising_results}, we present our findings for the Ising model $(J_2=0)$.
In this case, to guarantee that the topologically frustrated and non-frustrated models share the same correlation length, it is sufficient to consider systems with periodic boundary conditions (PBC), composed of the same odd number of spins, subject to the same value of $h$, but with opposite values of $J$ — specifically, $+1$ for the frustrated model and $-1$ for the non-frustrated one.
In the upper panel, we display the raw data obtained for the REC as a function of the applied magnetic field, both with and without TF. 
Although our data are derived from systems of finite sizes, two distinct behaviors corresponding to the two phases of the model are observable. 
For large magnetic fields $h$, the REC of the TF and non-TF Ising model converges to the same value. 
On the other hand, when the interaction term dominates over the external field, the behaviors differ significantly, and the difference increases with the system size.
In particular, the limit $h \rightarrow 0^+$ is of special interest in this region. 
For the unfrustrated system, as $h$ approaches zero, $C(\rho)$ becomes independent of $L$ and tends to 1, consistent with the fact that the ground state can be well approximated by the GHZ state $\ket{\phi}=\frac{1}{\sqrt{2}}(\ket{\uparrow}^{\otimes L} + \ket{\downarrow}^{\otimes L})$, where $\ket{\uparrow}$ and $\ket{\downarrow}$ are the eigenstates of $\sigma^z$ with eigenvalues $+1$ and $-1$, respectively. 
In contrast, with TF, for $h \to 0^+$ the REC tends to $1 + \log_2(L)$, consistently with what can be inferred from the expression for the ground state of the TF Ising model presented in the previous section.

Increasing $h$ but remaining in the same phase, the independence on the size disappears and the QC in the unfrustrated system shows a volume-law, which recalls the behavior of magic, as measured by the stabilizerness R\'{e}nyi entropy~\cite{Odavic2023}, as can be seen in the middle panel of Fig.~\ref{Ising_results}. 
The analogy between the stabilizer R\'{e}nyi entropy and the REC holds in TF systems as well, where both exhibit a volume-law behavior in the large-size limit, although with a significant correction for smaller systems.

The bottom panel of Fig.\ref{Ising_results} shows our numerical results for $R$ in the case of the Ising model as a function of the magnetic field. 
Consistent with our working hypothesis, we observe that, as $L$ diverges, the value approaches $1$ throughout the whole topological phase and zero in the paramagnetic phase.
The speed at which these values are reached depends on the ratio between the correlation length and the system size.
Since the correlation length diverges at the phase transition, finite-size effects vanish more rapidly the further one moves away from $h=1$.
We can then write the REC in the thermodynamic limit as the sum of the topological term and the magnetic one. 

\begin{figure}[t!]
    \centering
   \begin{subfigure}[t]{0.95\columnwidth}
        \centering        \includegraphics[width=0.95\columnwidth]{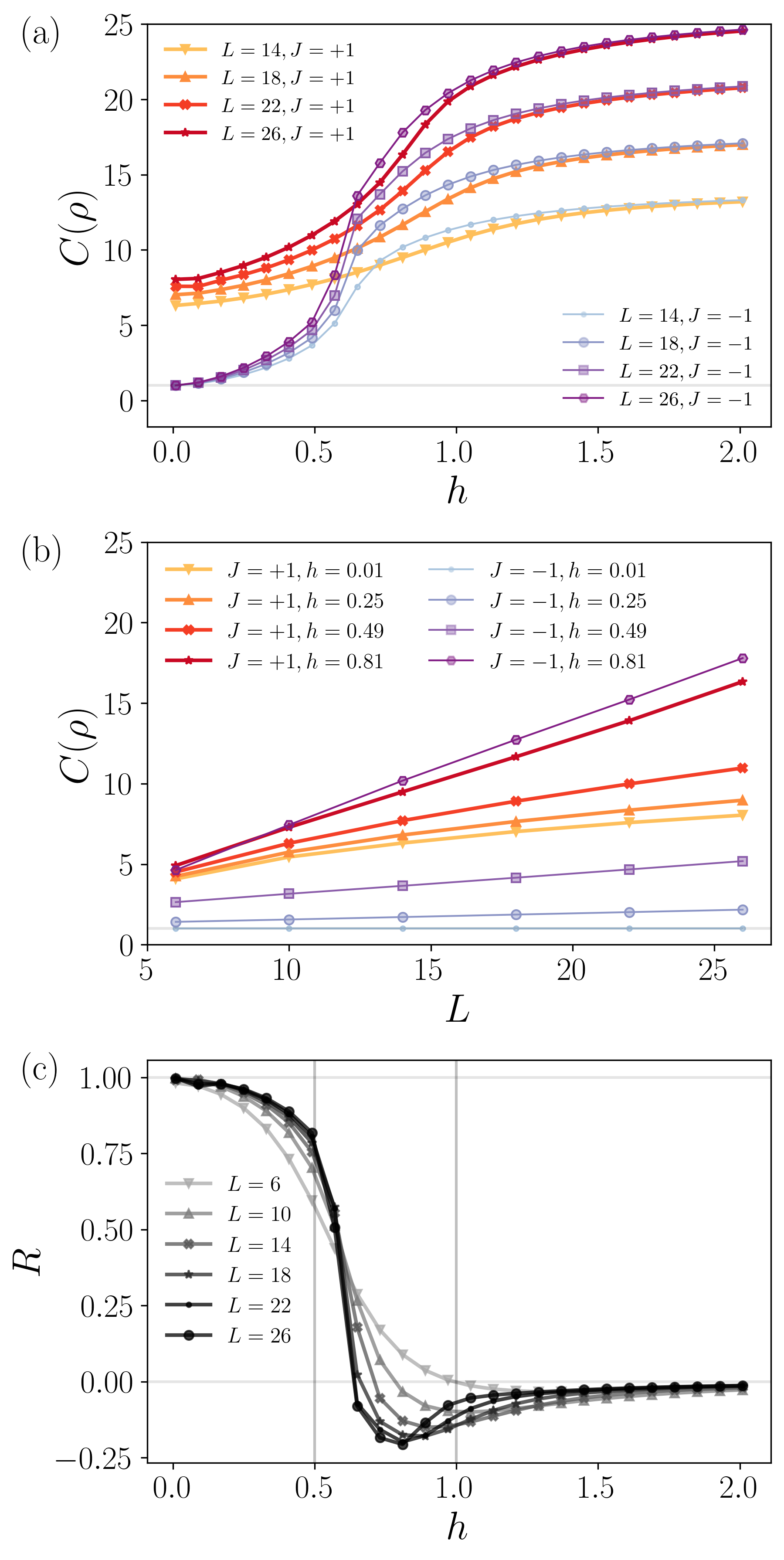}
    \end{subfigure}
    \caption{(a) Quantum Coherence of the ground state of the TF ANNNI chain (red full markers) and of the corresponding unfrustrated OBCs model (blue empty markers) as a function of the magnetic field and for different values of the system size. (b) The same quantities are plotted as a function of the system length for different values of the magnetic field. (c) The ratio $R$ in eq.~\eqref{eq:R_def} for the ANNNI model as a function of the magnetic field and for different system sizes.}
    \label{ANNNI_results}
\end{figure}

Let us now consider the ANNNI model by introducing an antiferromagnetic next-to-nearest-neighbor interaction ($J_2 > 0$).
Even if the two models seem not so different from each other, the ANNNI model is much more complex to analyze, and several technical challenges must be addressed. 
The first problem arises when trying to associate the specific non-TF model with its TF equivalent, i.e., when we try to select the two states (one frustrated and one unfrustrated) that share the same correlation length.
For the Ising model, this link is natural. 
Indeed, assuming periodic boundary conditions (PBCs) and an odd number of spins, transitioning between TF and non-TF Ising systems can be achieved simply by inverting the sign of $J_1$. 
This approach does not apply to ANNNI models. 
Specifically, when PBCs are imposed and the number of spins is even but not an integer multiple of 4, the presence of the TF is unaffected by the sign of $J_1$. 
On the other hand, by changing the sign of $J_2$, we not only remove TF but also the extensive frustration characteristic of the model, thereby fundamentally altering the properties of the system. 
Therefore, according to~\cite{torre2024interplaylocalnonlocalfrustration}, this comparison is done by opening up the ring and moving from PBCs to open boundary conditions (OBCs). 
This introduces finite-size effects in the unfrustrated system, which are expected to become irrelevant in the thermodynamic limit.

The results for REC in the ANNNI model are shown in Fig.~\ref{ANNNI_results}. 
In analogy with the Ising model, the presence of TF induces, in the limit $h \rightarrow 0^+$, a dependence of REC on $L$ that is directly related to the dimension of the ground state manifold at the classical point. 
In this case, it is possible to show (see Appendix~\ref{ANNNI_coherence}) that, in the limit $h \rightarrow 0^+$, the REC of the TF system becomes equal to \mbox{$C(\rho_{\textrm{fr}}^{h\rightarrow0^+}) = \log_2[L(L+2)]-\log_2(e)$}. 
On the contrary, without topological frustration, the system allows a ground state that is a homogeneous superposition of four elements of the computational basis, and hence $C(\rho_{\textrm{unfr}}^{h\rightarrow0^+})=2$.  
As $h$ increases, the difference between the REC for the TF and non-TF models decreases, until, in the floating phase, the REC for the non-TF model becomes larger than that for the TF model. 
This behavior appears to be a characteristic feature of the floating phase and could serve as a starting point for future investigations of this elusive phase of the model.

The bottom panel of Fig.\ref{ANNNI_results} shows the behavior of the ratio in~\eqref{eq:R_def} as a function of $h$ for a fixed value of $\kappa=|J_1|/J_2 > 1/2$ and allows us to test our working hypothesis.
As we can see, despite the relatively small size of the system considered, constrained by the large bond size in the DMRG algorithm required to obtain an accurate estimate of the ground state, the thermodynamic behavior is evident.
In the TF phase, the ratio Eq.~\eqref{eq:R_def} approaches 1 while, in the paramagnetic phase, it drops to 0 in a complete analogy with the Ising model.

\section{Conclusions and  outlook}\label{conclusion}

In summary, we examined quantum coherence in two one-dimensional models—the Ising and the ANNNI model—comparing its behavior in the presence and absence of topological frustration. 
In both cases, we found that in the paramagnetic phase, quantum coherence remains unaffected by this source of frustration in the thermodynamic limit. 
However, in the frustrated phase, topological frustration—though not the only source of frustration in the ANNNI model—introduces a correction that scales logarithmically with the system size.

Our findings confirm that, at least in one-dimensional models, topological frustration induces in quantum coherence the same structural behavior previously identified in entanglement and magic. 
This leads to a clear separation between two contributions: a local term dependent on the Hamiltonian parameters and a topological term that remains constant throughout the phase.

Leveraging its phase invariance, and drawing parallels with other quantum resources, we analytically derived the expression for the contribution associated with topological frustration using the explicit ground state near the classical point.
This separation is further emphasized by the distinct scaling behaviors of the two terms: for quantum coherence, as in magic~\cite{Odavic2023}, the local contribution scales extensively, while the topological component follows a logarithmic dependence on system size, reflecting the dimension of the ground state manifold at the classical point.
In contrast, for entanglement, the local term follows an area law, remaining independent of system size $L$, whereas the topological term depends on the logarithm of the ratio between the length of the subsystem $M$ and $L$.

At this point, a natural question arises: should the behavior observed for quantum coherence, entanglement, and magic be regarded as a universal feature of all quantum resources when evaluated on the ground states of topologically frustrated systems in the thermodynamic limit? 
While identifying multiple cases where this pattern holds does not constitute a formal proof, the fact that all analyzed quantum resources exhibit the same behavior strongly supports this conjecture.

Let us comment that a decomposition of any entropic resource into the sum of local and topological contributions would follow if it was possible to effectively write the ground state of a TF chain as $\ket{0}\bra{0} \otimes \ket{\rm kink}\bra{\rm kink}$, where $\ket{0}$ is the ground state of the non-frustrated system and $\ket{\rm kink}$ is the state containing the delocalized kink excitation(s) observed close to the classical point. 
Clearly, such an expression is mathematically ill-defined (for instance, the Hilbert space dimension is mismatched), but expresses the physical intuition that TF is able to inject into a system stable topological excitations which, in the thermodynamic limit, propagate over an unaffected, unfrustrated substrate. 
To make mathematical sense of this intuition, a possible avenue to prove the decomposition of resources could involve extending the result presented in~\cite{PhysRevA.93.012303} for mutual information, already used in \cite{10.21468/SciPostPhysCore.7.3.050} to prove the topological nature of the TF phase. 
In both cases, a crucial role is played by the possibility to link the Rényi entropy of order $2$, there used to quantify entanglement, to the swap operator. 
In this way, the (entropic) resource is written as the $\log$ of the trace of reduced density matrix times an operator. 
Such formulation allows one to invoke a (generalized) adiabatic evolution and proves that a decomposition of the resource into the (topological) contribution close to the classical point and local one due to the local deformation of the state reflecting the local correlations. 
To generalize such an approach, one would need to know that all entropic resources can be written as expectation values of some operator with respect to a density matrix and to prove that the adiabatic continuation and decomposition would apply with any such representation. 
We hope to be able to provide such construction in the future but have to admit that for now, rigorous and general proof of our hypothesis remains an open question.

The analysis carried out in this work was made possible by using the algorithm introduced in~\cite{kožić2025computingquantumresourcesusing} to calculate the relative entropy of coherence. 
This method leverages the efficient MPS representation of the system’s ground state in combination with the TCI algorithm class to perform measurements. 
By adopting this approach, it is possible to overcome the challenge of the exponential growth of the system’s Hilbert space, hence exploring larger system sizes.
Furthermore, this technique is sufficiently general to compute a wide range of observables and quantum resources, offering the potential to further investigate the properties of topologically frustrated systems in the thermodynamic limit, a direction we plan to pursue in future work.

\begin{acknowledgments}
SBK and GT thank Nora Reinić, Daniel Jaschke, and Simone Montangero for useful discussions and valuable insights. 
SBK acknowledges support from the Croatian Science Foundation (HRZZ) Projects DOK-2020-01-9938.

\end{acknowledgments}

\appendix

\section{Quantum Coherence of ANNNI ground state near the classical point}\label{ANNNI_coherence}

In this Appendix, we compute analytically the quantum coherence of the ground state of the TF ANNNI chain made of a number of spins that is even but not an integer multiple of four near the classical point.
To do this, we will exploit the ground state expression achieved in~\cite{torre2024interplaylocalnonlocalfrustration} using perturbation (and graph) theory valid for $h \ll 1$
\begin{equation}\label{ANNNI_GS}
    \ket{g} = A \sum_{k=1}^{L}\sum_{p=k}^{k+L/2-1} \!\! \sin\left[\frac{(p-k+1) L \pi}{L+2}\right] \ket{\psi(k,p)}.
\end{equation}
Here $A=2/\sqrt{L\left(L+2\right)}$ is the normalization constant, while  $\ket{\psi(k,p)}\!\equiv\!\ket{2k-1,(-1)^{k}} \ket{2p, (-1)^{p+1}}$ are orthonormal states made of the tensor product of two kink states. 
The first of the two kink states lives in the sublattices made by the even spins while the second in the orthogonal sublattice. 
The two indices in each kink state refer, respectively, to the position of the kink in the sublattice and to the sign of the kink where +1 (-1) stands for $\uparrow \uparrow$ ($\downarrow \downarrow$). 

\begin{figure}[t]
    \centering
    \begin{subfigure}[h]{0.90\columnwidth}
        \centering
    \includegraphics[width=0.90\columnwidth]{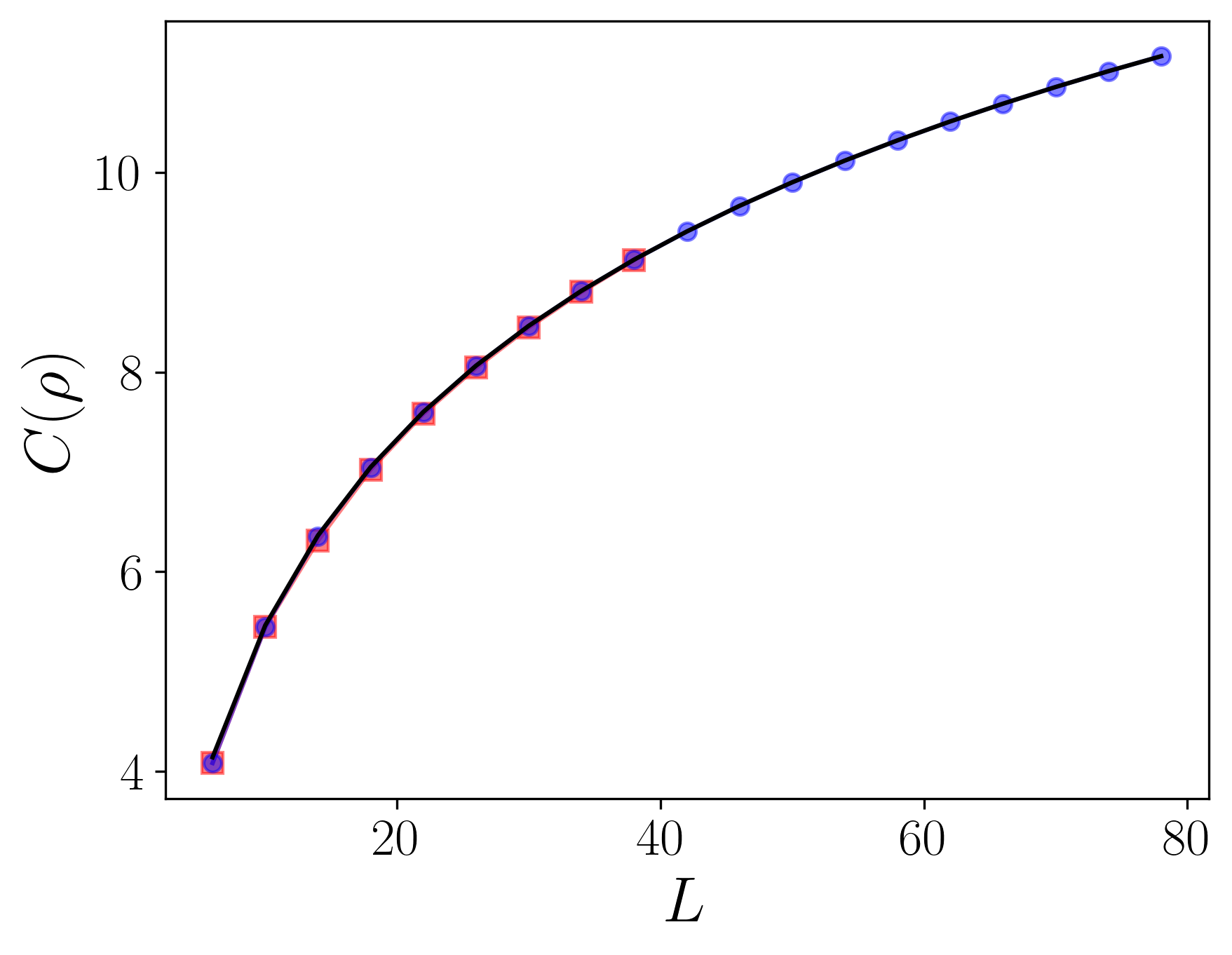}
    \end{subfigure}
    \caption{Comparison of the different results for the REC close to the classical point of the ANNNI model.
    The black line stands for the analytic expression in Eq.~\ref{analytic_ANNNI}.
    The blue circular points indicate the results obtained from the exact diagonalization of the ANNNI in the first-order perturbation theory ($h \rightarrow 0^+$).
    The red squares indicate the results of a numerical analysis that exploit 
    the procedure outlined in Sec.~\ref{DMRG_TCI_Section} with Hamiltonian parameters equal to $h=10^{-3}, \kappa=1.0$ and tolerance set equal to $\epsilon = 10^{-12}$
    }
    \label{ANNNI_analyt_num_comparison}
\end{figure}

Applying eq.~\eqref{QC_pure} to the state in eq.~\eqref{ANNNI_GS}, the quantum coherence becomes
\begin{eqnarray}\label{ANNNI_QC}
C(\rho_{\textrm{fr}}^{h\rightarrow0^+}) & = & -
    \frac4{L + 2}\sum_{r = 0}^{\frac L2 - 1}
    \sin^2\left[\alpha(r + 1)\right] \cross \nonumber \\
    & & ~~~\cross\log_2\left[\frac4{L(L + 2)}
        \sin^2\left[\alpha(r + 1)\right]\right]\text, ~~~~
\end{eqnarray}
where $\alpha = L\pi / (L + 2)$.
To obtain a closed form, we start splitting Eq.~\eqref{ANNNI_QC} in two terms separating the $L$ contribution in the logarithm from the $\sin$ function. 
The first term is given by
\begin{align*}
    & \frac{-4}{L + 2}\log_2\!\left[\frac4{L(L \!+\! 2)}\right]\!\!
    \sum_{r = 0}^{\frac L2 - 1}
    \sin^2\left[\alpha(r \!+\! 1)\right] = \log_2\!\left[\frac{L(L\! +\! 2)}{4}\right],
\end{align*}
where we have explicitly taken into account that the number of sites is even. The second term does not admit a simple closed form. To handle it we consider its thermodynamic limit ($L\rightarrow \infty$)
\begin{align*}
    &\frac{-4}{L + 2}\int_{-\frac12}^{\frac L2 + \frac12}\mathrm dx
    \sin^2\left[\alpha(x + 1)\right]
    \log_2\left[
    \sin^2\left[\alpha(x + 1)\right]
    \right] = \\
    &\frac{-4}{\ln(2)(L + 2)}\int_{-\frac12}^{\frac L2 + \frac12}\mathrm dx
    \sin^2\left[\alpha(x + 1)\right]
    \ln\left[
    \sin^2\left[\alpha(x + 1)\right]
    \right] = \\
    &= \frac{-4}{L + 2}\left(\frac{L + 2}2\right)\frac{1}{\ln(2)} \times \nonumber \\
    & \times \left(\frac1{\frac{\alpha(L + 2)}2}\int_{\frac\alpha2}^{\frac\alpha2(L + 3)}\mathrm dy \sin^2(y)\ln\left[\sin^2(y)\right]\right) \nonumber \\
     &= \frac{1}{\ln(2)}(2 \ln (2)-1)
\end{align*}
where we averaged the integral in parentheses over a long interval, which we can safely replace with an average over the period $ [0, 2\pi]$. Combining the two contributions,  we finally find in the $L\rightarrow \infty$ limit
\begin{equation}\label{analytic_ANNNI}
    C(\rho) = \log_2[L(L+2)]-\log_2(e).
\end{equation}
In Fig.~\ref{ANNNI_analyt_num_comparison}, we compare this formula with the numerical results obtained through the procedure outlined in Section~\ref{DMRG_TCI_Section}, finding good agreement even for small systems. 
To further support our claim, we also compare our results with the numerics derived from the exact expression for the ANNNI ground state, given by Eq.\eqref{ANNNI_GS}, near the classical point~\cite{torre2024interplaylocalnonlocalfrustration}.

\section{Basis Dependence of Quantum Coherence}\label{rotation}

It is well known that the value of quantum coherence (QC) is not solely determined by the quantum state itself, but also depends on the basis in which the state is expressed. 
This naturally raises the question of whether the decomposition of QC observed in topologically frustrated (TF) systems remains valid when the ground-state density matrix is expressed in a basis different from the computational one we used the bulk of this paper. 
In this appendix, we address this question by analyzing the Ising limit of the model, obtained by setting $J_2 = 0$ in Eq.~\eqref{eq:Hamiltonian}.

Given the central role of TF systems in our analysis, it is natural to classify possible basis transformations into two categories.
The first consists of transformations obtained from the computational basis through local unitary operations of the form:
\begin{equation}
\label{localU}
\mathcal{U} = \bigotimes_{i=1}^L U_i.
\end{equation}
These transformations preserve the spatial structure of the system and, by extension, its topological features.
The second category includes all other (non-local) transformations that do not conform to Eq.~\eqref{localU}, and which can alter the geometry of the system, potentially destabilizing its topological characteristics.

To maintain clarity and focus, we restrict our analysis in this appendix to the first category. 
Additionally, we assume all local unitaries are identical and given by:
\begin{equation}
\label{localUI}
U_i \equiv \cos\theta\,\mathbb{1}_i + \imath\,\sin\theta\,\sigma^y_i,
\end{equation}
here $\theta$ is a continuously tunable parameter.
Although Eq.~\eqref{localUI} does not represent the most general local rotation, it suffices to demonstrate the robustness of our conclusions.
It is important to note that for $\theta=n\pi/2$ (with $n\in\mathbb{N}$) the resulting basis is equivalent to the computational basis, modulo a permutation of its vectors. 
More generally, $\theta$ and $\theta'=\theta+n\pi/2$ (with $n\in\mathbb{N}$) define the same basis, differing only by a reordering of the basis elements.
Therefore, we will limit our analysis in the interval $\theta\in[0,\pi/2]$.

We begin our analysis by examining how such local rotations affect the QC near the classical point, i.e., for $h \rightarrow 0^+$.
Recall that in this limit, in absence of topological frustration, the ground state becomes a GHZ state, while with frustration, it forms an equal superposition of all kink states.
In the computational basis, these two states exhibit markedly different QC: the GHZ state yields a constant QC value of 1, independent of the system size $L$ whereas the frustrated superposition leads to QC scaling as $\log_2 L + 1$.

\begin{figure}[H]
	\centering
	\begin{subfigure}[t!]{0.88\columnwidth}
		\centering        \includegraphics[width=0.88\columnwidth]{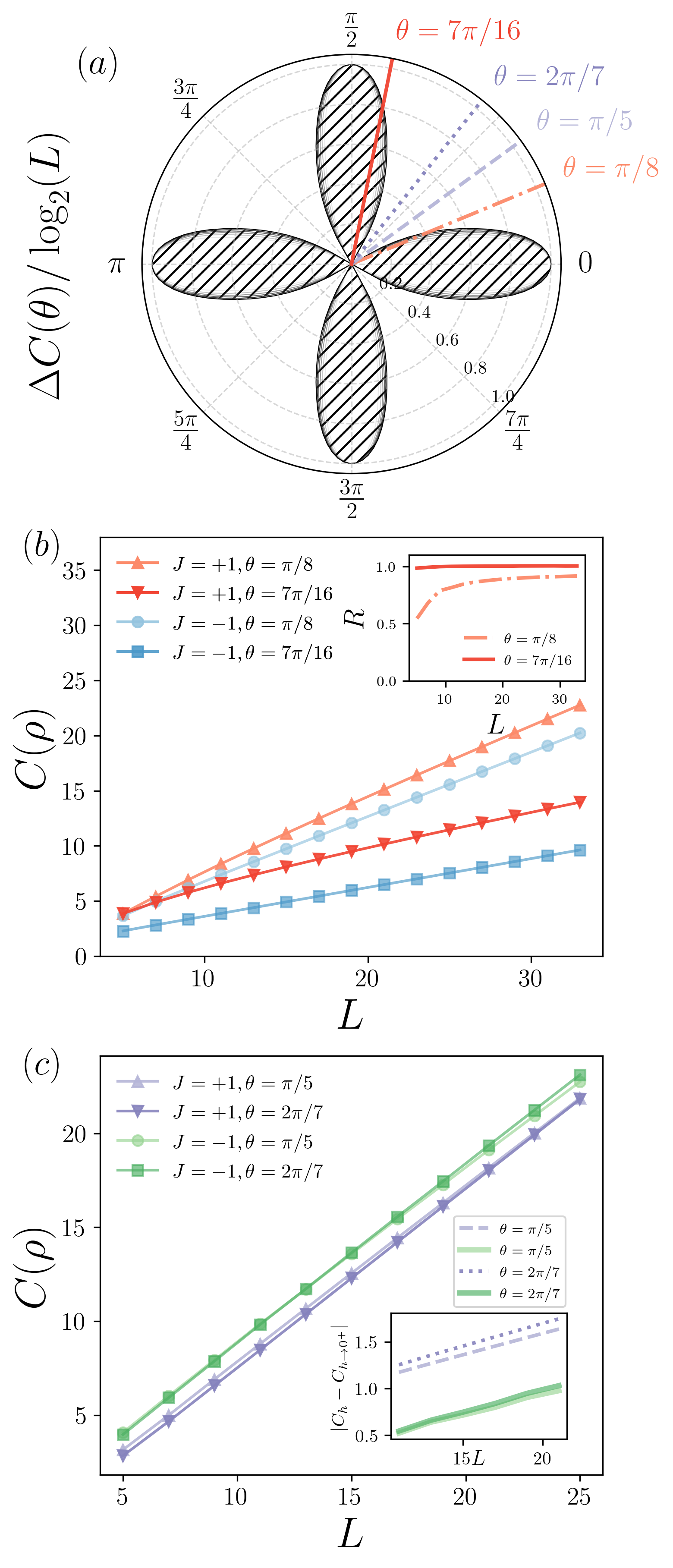}
	\end{subfigure}
	\caption{(a) Angular plot of the topological contribution to REC from Eq.~\eqref{DeltaCtheta}, as a function of the rotation angle $\theta$. Shaded areas represent the range of validity for the ratio in Eq.~\eqref{eq:R_def}, that is with a finite denominator: at the classical point $h\rightarrow 0^+$(i.e. $h=0.001$) the denominator vanishes for angles between $\pi/6 \lesssim \theta \lesssim\pi/3$. 
	(b) QC as function of $L$ for the Ising chain with $h=0.5$, $J=\pm1$ (red and blue respectively), and $\theta = \pi/8, 7\pi/16$.
	Inset is the corresponding ratio $R$ from Eq.~\eqref{eq:R_def} for the same angles $\theta = \pi/8,7\pi/16$. 
	(c) QC as function of $L$ for the Ising chain with $h=0.5$ and $J=\pm1$ (purple and green respectively) for angles $\theta=\pi/5,2\pi/7$. 
	Inset is the corresponding value of  $|C_{h=0.5} - C_{h=0.001}|$ for $J=\pm1$ (purple and green curves respectively) for the same angles.}
	\label{fig:coherence_rot}
\end{figure}

Upon rotating the basis using Eq.~\eqref{localUI}, numerical analysis reveals a consistent behavior: for all the bases different from the computational one, independently on the presence of topological frustration, QC shows a leading term proportional to $L$.
Nevertheless the difference:
\begin{equation}
\Delta C(\theta) = \left| C(\rho_{\textrm{fr}}^{h \rightarrow 0^+}) - C(\rho_{\textrm{unfr}}^{h \rightarrow 0^+}) \right|,
\label{DeltaCtheta}
\end{equation}
that appears in the denominator of Eq.~\eqref{eq:R_def}, is governed solely by subleading contributions, at most scaling as $\log_2 L$, as shown in panel (a) of Fig.~\ref{fig:coherence_rot}.
The figure reveals two distinct regimes in the behavior of $\Delta C(\theta)$  as a function of lattice size $L$. 
In the first regime, ranging approximately over $0\le\theta\lesssim \frac{\pi}{6} \cup \frac{\pi}{3}\lesssim\theta\le\frac{\pi}{2}$,
the difference scales as $\log_2 L$. Outside this region, however, the ratio $\Delta C(\theta)/\log_2 L$ tends to zero while $L$ increases. 

Because the two regimes differ qualitatively, we analyze them separately.
In the first regime, $\Delta C(\theta)$  scales logarithmically with $L$, just as it does in the computational basis. This indicates that the decomposition of QC into local and topological components holds throughout this entire region. 
Panel (b) of Fig.~\ref{fig:coherence_rot} supports this conclusion, with a finite size scaling performed for two representative angles, yielding the same behavior for the ratio in Eq.~\eqref{eq:R_def} we observed in the computational basis.
In the second regime $\frac{\pi}{6}\lesssim\theta\lesssim\frac{\pi}{3}$, the picture becomes more subtle. 
Here, not only does $\Delta C(\theta)$ lack a logarithmic contribution, but individual QC values for both frustrated and unfrustrated states also lack any scaling with $\log_2(L)$ as shown in panel (c) of Fig.~\ref{fig:coherence_rot}. 
This complicates the separation of topological and local contributions, as both are dominated by linear terms and show subleading corrections that vanish or remain constant in the thermodynamic limit.
Moreover, as detailed in App.~\ref{Bond}, numerical computations in this regime become increasingly challenging.
To disentangle local from topological contributions under these conditions, we compare the scaling slopes of QC for different values of the transverse field $h$. 
Since the local contribution to QC depends on $h$ and vanishes as $h \rightarrow 0^+$,  while the topological component remains constant up to the critical point, the local term can be isolated by subtracting the QC at finite $ h$  from its value in the classical limit.
This approach is applicable to both frustrated and unfrustrated systems.
If the local term is correctly isolated, its thermodynamic limit should match in both cases. 
This means that, since such a difference scales with $L$ their slope must be equal, 
which is indeed what we observe and can be seen in panel (c) and the accompanying data table~\ref{tab:best} reporting the best-fit parameters.

\begin{table}[t!]
  \centering
  \begin{tabular}{ |c|c|c| } 
    \hline
    $\theta$ & $\delta C$ for $J=-1$ & $\delta C$ for $J=+1$ \\ 
    \hline
    $\pi/5$   & $0.045 L+0.047$       & $0.046 L+0.670$       \\ 
    $2\pi/7$  & $0.049 L+0.026$       & $0.048 L+0.722$       \\ 
    \hline
  \end{tabular}
  \caption{Best‐fit parameters of the difference 
    $\delta C=|C_{h=0.5}-C_{h=0.001}|$ as a function of the system size $L$ 
    for both frustrated ($J=-1$) and unfrustrated ($J=+1$) ground states 
    at two representative angles in the interval 
    $\frac{\pi}{6}\lesssim\theta\lesssim\frac{\pi}{3}$.}
  \label{tab:best}
\end{table}

Thus, even in this second regime, we can conclude that the QC in TF systems decomposes into two contributions: one local and one topological in nature.
Before concluding this appendix, we highlight an interesting observation: in the cases we analyzed, the inclusion of the local term reduces the total QC instead of increasing it.
Although counterintuitive at first glance, this effect can be understood by noting that local rotations such as those considered can greatly enhance QC.
For instance, consider the product state $\ket{\uparrow}^{\otimes L}$, which has zero QC in the computational basis but reaches the theoretical maximum REC value of $L$ for $\theta=\pi/4$.
Since this maximum corresponds to the largest QC achievable for an 
$L$-qubit system, any introduction of correlations—such as those due to frustration—can only decrease the QC from this saturated value.

\section{Details about numerical analysis}
\label{Bond}


In our work, we made a large use of numerical evaluations, all of them based on an MPS representation of the ground state and the corresponding MPS generated by the TCI algorithm~\cite{TensorCrossInterpolation.jl}.
In fact, although topological frustration induces non-local correlations in a system whose entanglement entropy otherwise obeys an area law, the ground state can still be represented as a matrix product state~\cite{Torre2024}. 

As for the DMRG algorithm that provides the ground state of the system, this has been adapted to the particular requirements of the systems under analysis.
To handle the requirement of periodic boundary conditions (PBCs) (for which it is known that periodic DMRG is not as efficient as the standard OBC case~\cite{Verstraete_2004}), we add an extra long-range correlation inside the MPO structure following~\cite{Weyrauch_2013}.
An additional problem we had to face was the presence of energy gaps that close exponentially with the system size in the ordered phases of non-frustrated systems.
To solve such a problem, we use a parity projector $P_{\pm}=1/2\,\bigl(I \pm \hat{\Pi}_x\bigr)$ with $\hat{\Pi}_x$ being the parity operator along $x$, to ensure that the resulting state is not an unwanted superposition of the two states with minimal energy in the opposite parity sector. 
Extra care is also taken to ensure convergence near the classical point $h\rightarrow0^{+}$, as the variational procedure can stall in the presence of near-degeneracy, for this reason we construct the initial MPS using the known analytical expression of the ground state in this limit, namely GHZ state for the ferromagnetic case ($J=-1$) and the kink state in the antiferromagnetic case ($J=+1$).
Estimation of QC comes in handy as one knows exactly what the true value of the ground state in this limit should be (for example in the Ising case as already mentioned either $1$ or $\log_{2}(L)+1$), which is a more precise measure of the quality of the state rather than just using energy as a figure of merit. 
Once we applied these considerations to the illustrated problems, we were able to proceed with our simulations.
For all DMRG calculations, ITensor Library was used \cite{10.21468/SciPostPhysCodeb.4} with 50 sweeps and maximum bond dimension capped at $\chi_{\mathrm{max}}\equiv200$, with resulting MPS not exceeding $\chi_{\textrm{unfr}}=100,\chi_{\textrm{fr}}=150$ near critical point $h=1$ for $J=\mp1$ respectively.

\begin{figure}[t!]
	\centering
	\begin{subfigure}[t]{0.9\columnwidth}
		\centering        \includegraphics[width=0.9\columnwidth]{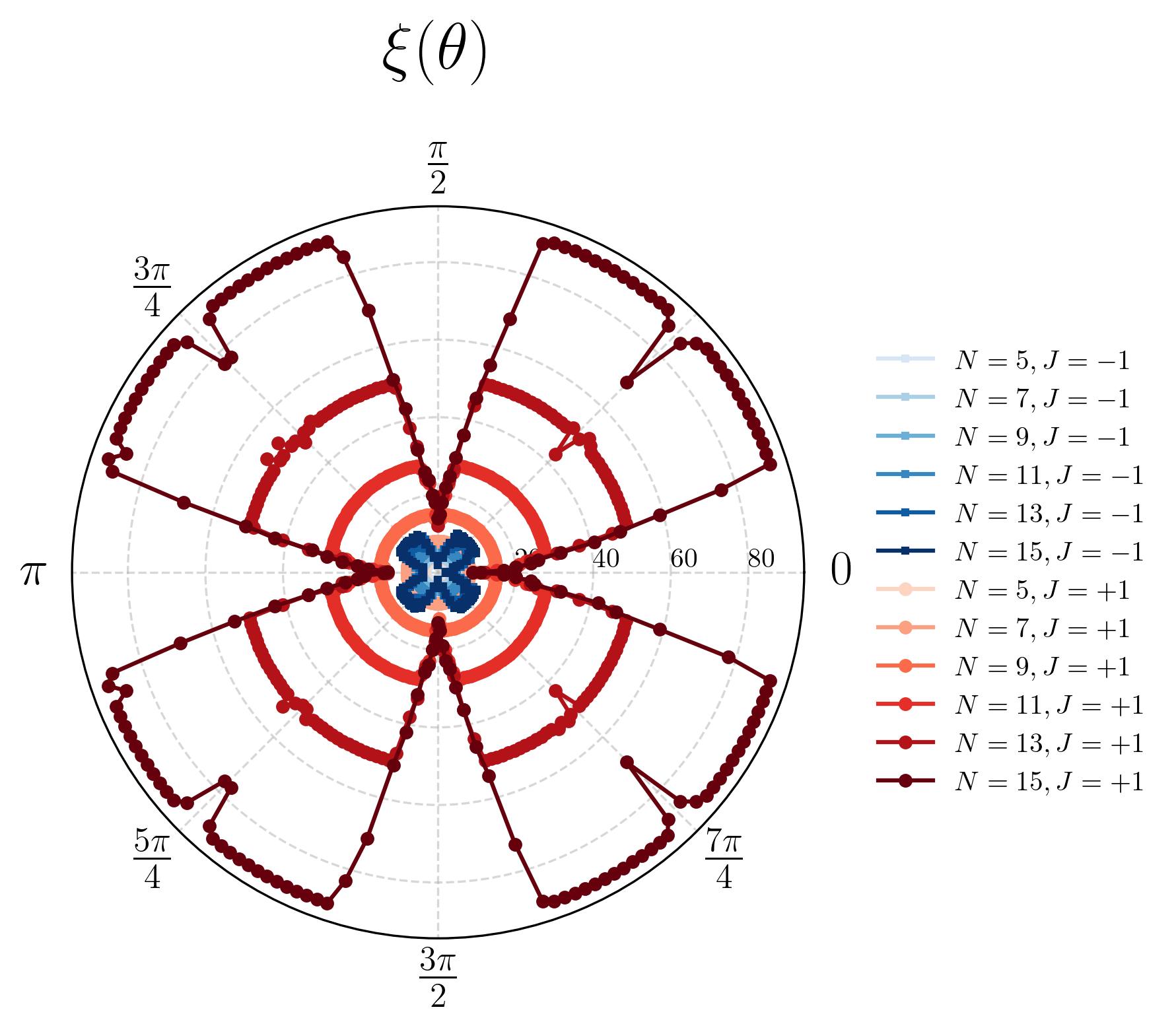}
	\end{subfigure}
	\caption{Maximum bond dimension $\xi$ of the TCI MPS required to reach precision $\epsilon=10^{-6}$ for coherence $C(\rho)$ at the classical point $h\rightarrow 0^+$ for $J=\pm1$ as a function of rotation angle $\theta$ for system sizes $L$ running from 5 to 15.}
	\label{fig:coherence_bond}  
\end{figure}

However, the most relevant computational cost then stems from the bond dimension required to achieve a given level of accuracy in the MPS produced by the TCI algorithm~\cite{fernandez2024}. 
Our procedure for computation of QC exhibits a computational complexity of $\mathcal{O}(2\mathcal{L}^2\xi^2\chi^2)$, where $\chi$ and $\xi$ denote the bond dimensions of the ground state MPS and of the functional MPS obtained via the TCI algorithm, respectively, and $\mathcal{L}$ is the system size~\cite{kožić2025computingquantumresourcesusing}.
The bond dimension of the MPS generated by the TCI algorithm is connected to the underlying sampling procedure. Within a specified tolerance, the system reliably converges to the desired result after a sufficient number of function calls. 
Increasing the precision (i.e., reducing the tolerance) leads to a larger number of function calls, thereby increasing the computational time required to reach convergence~\cite{Dolgov2020}. 
From a practical standpoint, very high precision tolerance $\epsilon=10^{-12}$ and $\xi_{\mathrm{max}}=200$ or more is required, particularly for the reliable computation of the ratio in Eq.~\eqref{eq:R_def}.
Another highlight of the TCI algorithm is that it can estimate the local tensor error thereby giving the user the information about how good or bad is the local approximation of the measured quantity. 
This provides the relevant information on whether the allowed maximum bond dimension should be increased and if the tensor network representation is efficiently compressible or not due to bond dimension bottleneck.

The estimation of QC becomes even more challenging in the case of its evaluation in the rotated bases made in App.~\ref{rotation}. 
Even though the bond dimension $\chi$ of the MPS ground state representation in the new basis remains fixed, since entanglement is not affected by the local rotations, the bond dimension $\xi$ of the functional MPS tends to increase rapidly due to the exponential number of terms involved in the superposition that defines the rotated state. 
This can be seen even in the classical limit $h\rightarrow0^{+}$ as shown in Fig.~\ref{fig:coherence_bond} for Ising model.
The bond dimension grows prohibitively, reaching $\xi>80$ for a small system as $L=15$ in the frustrated case $J=+1$. Notice that this explosion of the required bond dimension occurs in the region where the denominator of the ration R in Eq.~\eqref{eq:R_def} vanishes and even the topological contribution to QC becomes extensive. This numerical issue signals the underlying physics and explains why it is hard to correctly capture the thermodynamic limit behavior of the systems in this region.
 
Overall, this discussion highlights that although the MPS representation effectively captures the entanglement structure, it is not well-suited for representing non-linear functions of quantum states that cannot be efficiently approximated within a low-dimensional effective Hilbert subspace.
We expect that the same bottleneck will occur in the case of stabilizer R\'enyi entropy. 
Hence it will require great computational power in order to estimate such entropy up to good precision threshold, which itself presents an interesting challenge.

\bibliography{apssamp}

\end{document}